# Controllable growth of Al nanorods for inexpensive and degradation-resistant surface enhanced Raman spectroscopy

Stephen P. Stagon[1], Xuefei Tan[2], Yongmin Liu[2,3] and Hanchen Huang[1,2] *

[1] Mechanical Engineering, University of Connecticut, Storrs, CT 06269, USA

[2] Mechanical and Industrial Engineering, Northeastern University, Boston, MA 02115, USA

[3] Electrical and Computer Engineering, Northeastern University, Boston, MA 02115, USA

\* **Correspondence and requests** for materials should be addressed to H.C.H. (e-mail: h.huang@neu.edu)

**Surface enhanced Raman spectroscopy (SERS) has the capacity of detecting trace amount of biological or chemical matter, even single molecules, through the use of metallic nanostructures such as nanorods. Silver (Ag) and gold (Au) nanorods have led to the impressive enhancement of Raman signals, but they are either expensive, degrade fast over time, or suffer from poor sample repeatability. In contrast, Al is much less expensive, and Al nanorods could potentially be resistant to degradation over time due to the protection from native aluminum-oxide layers. Unfortunately, the controllable growth of Al nanorods has not been reported so far. This Letter reports, *for the first time*, the controllable growth of Al nanorods using physical vapor deposition (PVD); through the use of oxygen (O) surfactants. The enhancement factor of the Al nanorods in SERS is as high as 1250, and shows nearly no degradation after storage in ambient for 30 days or annealing at 475 K for one day.**

Raman spectroscopy is widely used to detect the characteristic vibrational and rotational modes of a molecule. Although Raman signals are intrinsically weak, they can be significantly magnified by metallic nanostructures through the excitation of surface plasmons [1]. In the SERS detection of biological and chemical compounds, the enhancement factor of Ag nanorods has reached nearly $\sim 10^9$ [2, 3, 4], and that of Au nanorods has reached $\sim 10^5$ [5, 6]. These noble metals are expensive, and are often assumed to be stable against degradation. Despite the nobility, the capability of the Ag nanorods to enhance Raman signals quickly degrades to half of the initial value in one day at room temperature, and decreases by over an order of magnitude over 30 days [7]. When elevated temperatures are involved, the nanorods coarsen even faster, as we have shown recently [8]. As a result, the degradation of the enhancement factor can be exacerbated at higher temperatures, unless complex surface modification is used [9].

It will be highly desirable if metallic nanorods are resistant to degradation, and it will be doubly beneficial if these nanorods are inexpensive enough for practical biochemical sensing applications. Al nanorods can potentially bring this double benefit – the stable native oxide layer on Al nanorods, although only a few nanometers thick, can likely prevent them from fast coarsening or degradation, and Al is inexpensive when compared to Ag or Au. To reap the double benefits, we must first be able to grow Al nanorods, and to grow them controllably.

In this letter we present the first controllable growth of inexpensive Al nanorods using PVD and then demonstrate degradation-resistant SERS. The controllable growth is made possible based on the insight from the recently developed theory of nanorod growth, and the theory-guided use of surfactants. According to the theory [10, 11, 12, 13], the growth of nanorods relies on multiple-layer surface steps – they must impose a large diffusion kinetic barrier to adatoms and must be kinetically stable. For Al, the multiple-layer surface steps impose only a small diffusion barrier of about 0.13 eV, according to quantum mechanics calculations [14]. In contrast, this barrier is 0.40 eV for Cu [15] and 0.56 eV for Au [10]. As a result, the growth of pure Al nanorods is *impossible* according to the theory of nanorod growth. This theoretical impossibility may be the reason of why controllable growth of Al nanorods has not been reported so far.

Beyond the impossibility of growing pure Al nanorods, the theory also reveals that growth of Al nanorods is both possible and controllable if one can modify the multiple-layer surface step diffusion barrier, for example, through the use of surfactants. Surfactants can modify the local diffusion environment and therefore change the diffusion barrier. That is, it is possible to increase the adatom diffusion barrier above 0.13 eV on Al surfaces through the introduction of surfactants. In fact, surfactants have been used in the PVD growth of thin films, as well as growth of nanostructures in chemical vapor, as a catalyst, and even in solution syntheses [16, 17, 18, 19, 20, 21, 22]. In PVD processes, oxygen (O) has been known to act as surfactant in the epitaxial growth of metallic thin films for over two decades [16, 22]. During epitaxy, O atoms float on metal surfaces and tailor the growth mode of the film through modifying local kinetic barriers [16, 22]. Therefore, we anticipate that O atoms could potentially modify diffusion kinetic barriers during nanorod growth. In this regard, we note two salient features of O atoms. First, O atoms are surfactants, and therefore they tend to float on surfaces. Second, aluminum oxides such as $Al_2O_3$ have stronger bonding than Al; consequently, the diffusion barrier of Al adatoms near the oxide is likely larger than near pure Al. Based on these two features, we propose to use the amount of O surfactant to control the diameter in growing Al nanorods.

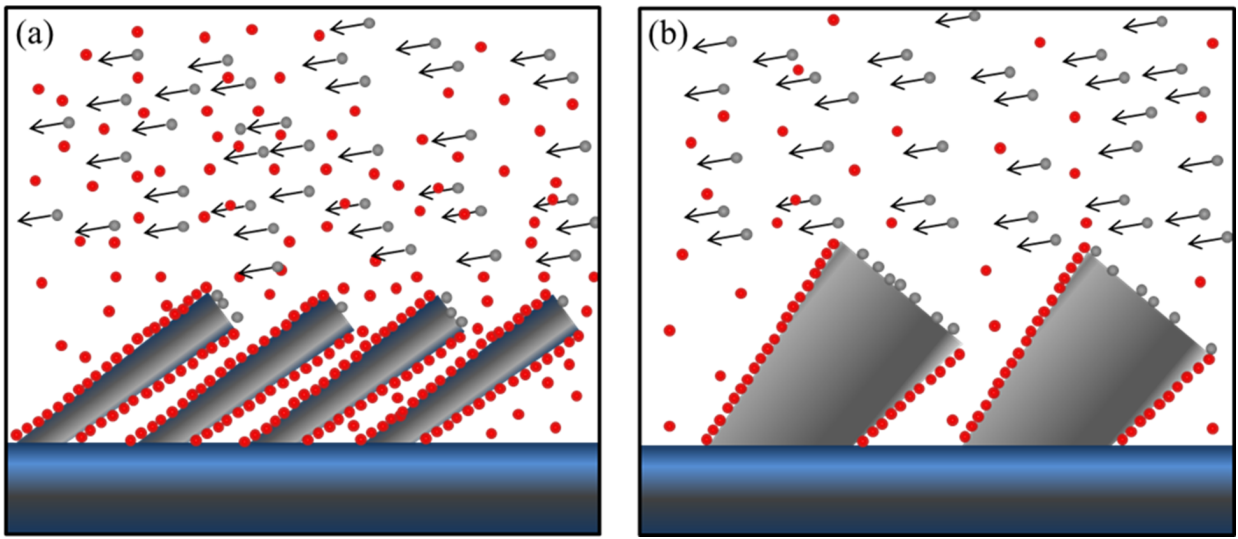

**Figure 1: Proposal of controllable growth of Al nanorods**. Schematic of growing Al nanorods (in gray), with O atoms (red spheres) as surfactants in PVD, at (a) a low vacuum (~$10^{-2}$ Pa for example) and (b) a high vacuum (~$10^{-5}$ Pa).

Figure 1 schematically illustrates our proposal of controllably growing Al nanorods. The PVD growth of nanorods uses glancing angle incidence. As a result, incident atoms land primarily on the top of nanorods. When the adatoms cannot easily diffuse from the top of nanorods to their sides, the nanorod diameter remains small. At low vacuum, there are enough O atoms in the PVD chamber to nearly completely decorate the sides of Al nanorods, so the diffusion of adatoms is slow (since the diffusion barrier is large) and the diameter is small. At high vacuum, there are too few O atoms to completely decorate the sides of Al nanorods, so diffusion of adatoms is not as slow and the diameter is not as small. At ultra-high vacuum, the diameter will become so large that continuous films, instead of nanorods, develop. In retrospective, one publication [23] that reports Al nanorods used relatively low vacuum (~$10^{-3}$ Pa). In addition, to grow Cu nanorods of small diameter, vacuum was intentionally interrupted to introduce oxide formation in another report [24]. Both of these two works serve as evidence that the proposal schematically shown in Figure 1 is physically reasonable.

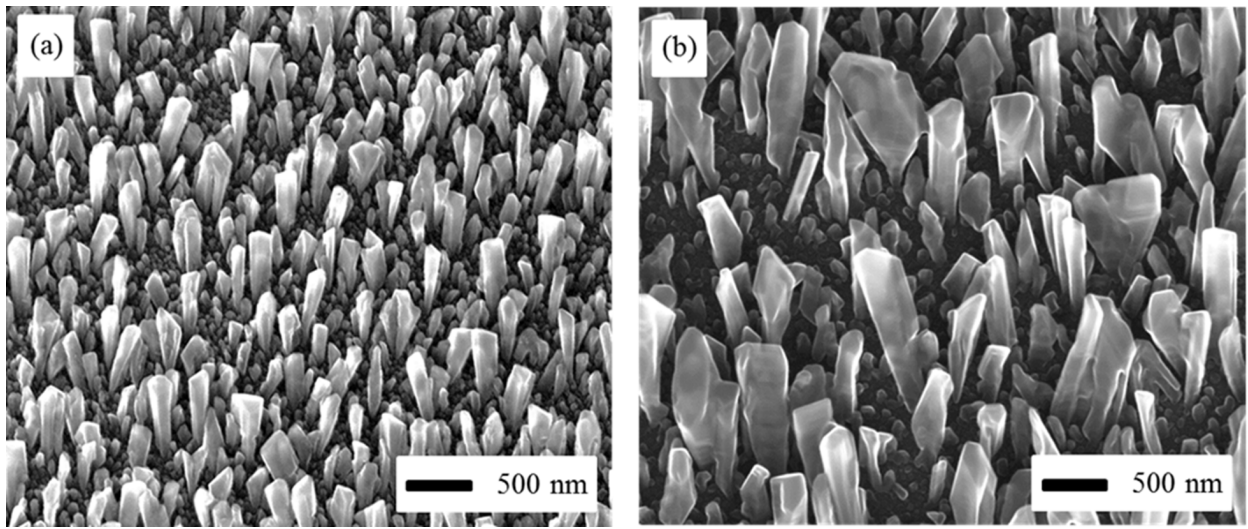

**Figure 2: Experimental realization of the proposal in Figure 1**. Scanning electron microscopy (SEM) images of Al nanorods grown at (a) a low vacuum of $10^{-2}$ Pa and (b) a high vacuum of $10^{-5}$ Pa.

Taking the proposal to experimental realization, we grow Al nanorods on a Si{100} substrate at two different vacuum levels. All other deposition conditions are the same; the substrate temperature is maintained at 300 K, the nominal deposition rate is 1.0 nm/s, and the incidence angle is 86°. As shown in Figure 2, lower vacuum leads to a smaller diameter of nanorods, about 125 nm; and the higher vacuum leads to a larger diameter of nanorods, as large as 500 nm. This set of results serves as experimental realization of the proposal in Figure 1 – the controllable growth of Al nanorods through the use of O surfactants. We recognize that nitrogen (N) concentration is also high. However, N loses to O in the reaction with Al; as the energy dispersive X-ray spectroscopy shows for Al nanorods that are as grown and annealed [Supplemental Material]. That is, O atoms, instead of N atoms, are the most likely surfactants during Al nanrood growth.

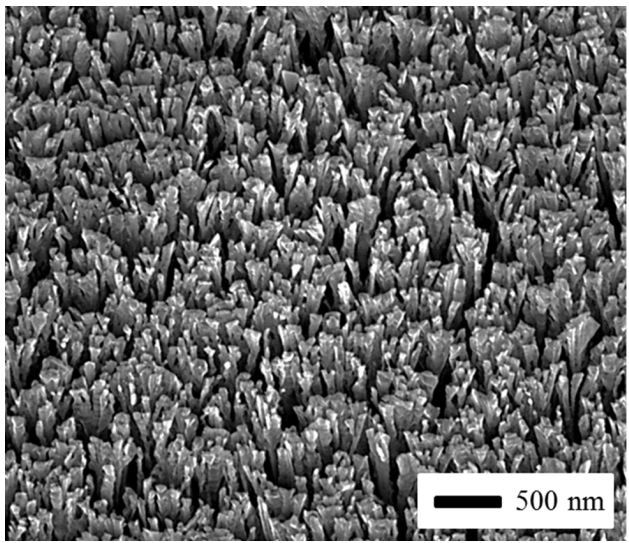

**Figure 3 Further evidence supporting the proposal in Figure 1**. SEM images of Al nanorods grown at a substrate temperature of 225K under a low vacuum of $10^{-2}$ Pa.

To provide further evidence, we also vary the substrate temperature, keeping the vacuum low at $10^{-2}$ Pa. At a lower substrate temperature of 225 K, a larger diffusion barrier in the presence of O surfactant should lead to even smaller diameter. Indeed, Figure 3 shows that the diameter of Al nanorods can be reduced to 50 nm.

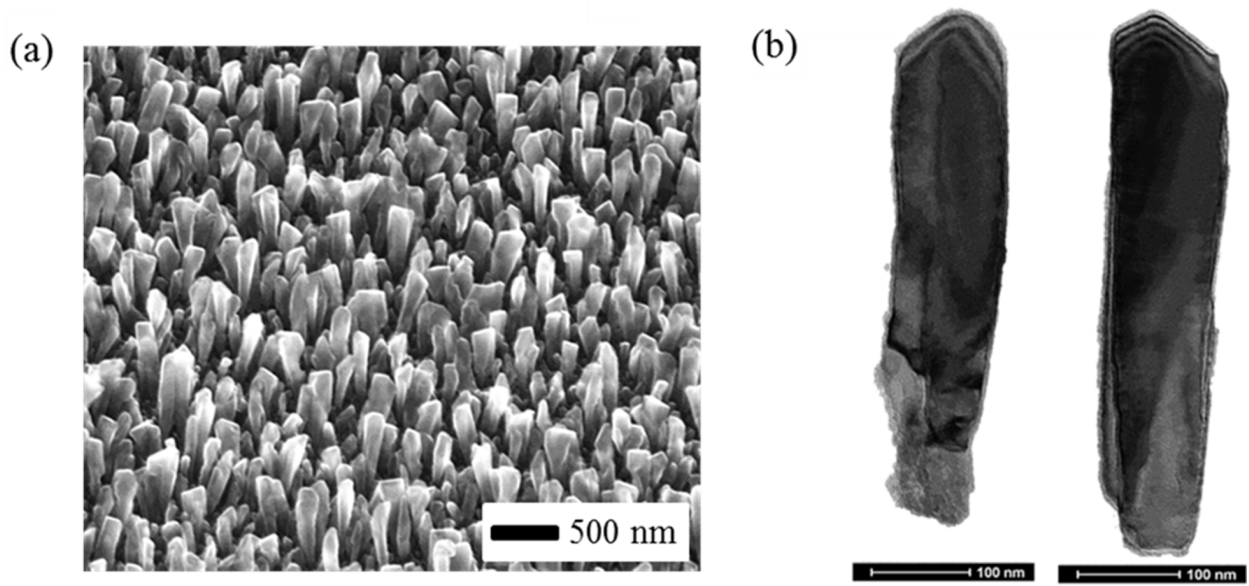

**Figure 4: Thermal stability of Al nanorods.** (a) SEM image of Al nanorods after annealing, and (b) TEM images of Al nanorods before (left) and after (right) the annealing.

Having established the controllable growth of Al nanorods using O surfactant, we also note that they are nearly uniform in diameter – the uniformity is an advantage for SERS. Beyond low cost, we also expect the Al nanorods to be stable against aging, due to the intrinsic oxide layer on surfaces. After annealing at 300 K in air (in ambient) for 30 days, the morphology of Al nanorods show no visible changes; image not included here. To further show this stability, we also annealed the Al nanorods of Figure 2a in air at 475 K (or 50% of Al melting temperature) for one day. As shown in Figure 4a, the morphology exhibits no visible change in comparison to that of Figure 2a. The transmission electron microscopy (TEM) images reveal that the Al nanorods are in the form of core-shell, with an oxide shell of about 5 nm in thickness. As Figure 4b shows, this thickness of 5 nm does not change by a visible amount after the 475 K annealing.

That is, the Al nanorods are stable both in time and at high temperature, and their structure and morphology are degradation resistant. In passing, we note that the core-shell structure is morphologically stable, although the structure changes through the growth of the oxide layer, up to 875 K; and that annealing to 1275 K results in a crystalline $Al_2O_3$ nanorod of the same dimension [Supplemental Material].

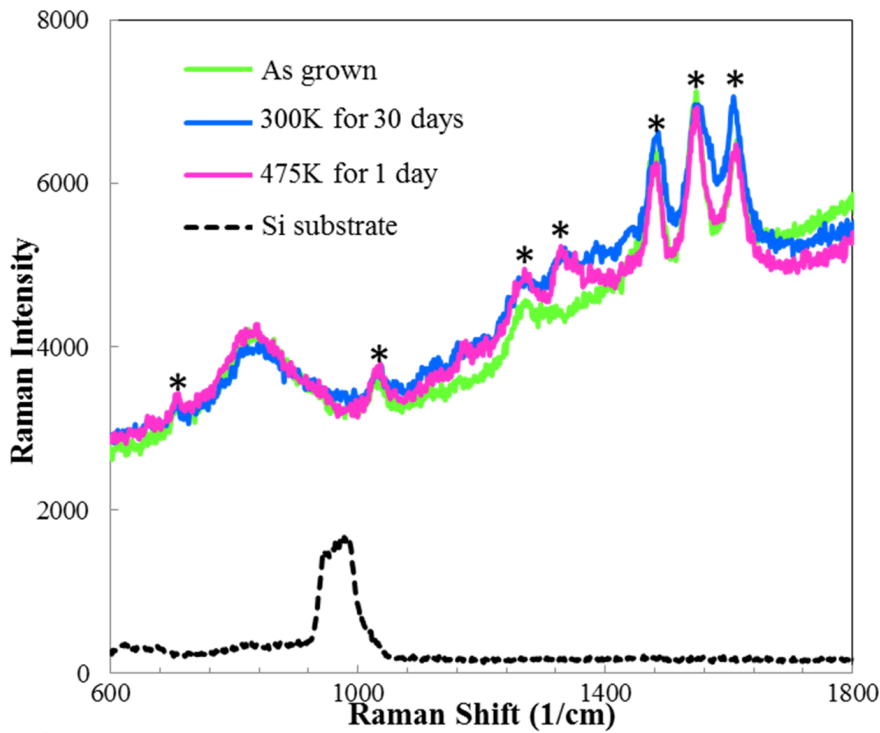

**Figure 5: Degradation-resistant SERS using Al nanorods.** Raman spectra of three sets of Al nanorods that are: as grown (green line), annealed at 300 K for 30 days (blue line), and annealed at 475 K for one day (pink line), respectively. The spectrum of Si{100} with 1 x $10^{-4}$ M N719 (dashed black line) is included for comparison, and asterisks mark the N719 peaks [25].

Having established the controllable growth and degradation resistance of Al nanorods, we now demonstrate their application, and advantages, in SERS. For this purpose, we use the Al nanorods realized at low vacuum and room temperature due to the simplicity of process, and use

Di-tetrabutylammonium-cis-bis(isothiocyanato)bis(2,2'-bipyridyl-4,4'-dicarboxylato) ruthenium (II) (N719) (Sigma Aldrich) as the chemical probe for comparison with literature results [24]. The Raman spectra are obtained using a 50 mWatt laser, with a 514.5 nm wavelength, and using an integration time of 10 seconds. The Raman spectra in Figure 5 reveal a salient feature. That is, SERS using Al nanorods is degradation resistant; there is little degradation after annealing at 300 K in air (in ambient) for 30 days or at 475 K in air for one day. Here, the data of Figure 5 is presented without any post processing to demonstrate the resistance to degradation. The degradation resistance is of particular merit when compared with the fast degradation of Ag nanorods based SERS. For Ag nanorods, the enhancement factor decreases by over 80% after annealing for 30 days at in ambient [7]. Due to this unique degradation resistance, Al nanorods are a candidate for SERS in applications that require long time stability or operate in harsh environments, where nanorods of other materials may degrade too much. In passing, we note that when the integration time is increased to 120 seconds, from the 10 seconds in Figure 5, the enhancement factor of the Raman peak at 1270 cm$^{-1}$ reaches about 1250 [Supplemental Material].

Before closing, we briefly discuss the repeatability of SERS measurements and the separation of Al nanorods from the substrate. First, repeatability of Al nanorods based SERS is high, with a standard deviation of only 6% over 15 SERS spectra that are randomly taken across the substrate of 7.5 cm in diameter. This standard deviation is comparable to that of Ag nanorods [26]. Second, it may be desirable to separate the Al nanorods from their substrate so that they may conformably coat any surface, such as in SHINERS [27]. We have separated Al nanorods from their substrate through simple vibration in a sonic bath. These separated Al nanorods remain in solution without aggregation or reaction due to their stable oxide shells.

To summarize, we report the controllable growth of Al nanorods using PVD, for the first time; through the use of O atoms as surfactants. Further, we demonstrate that both structure and morphology of Al nanorods are resistant to degradation with time, even at elevated temperatures up to 875 K. The SERS using Al nanorods is less expensive, and very importantly, more degradation resistant than that based on Ag or Au nanorods. The new mechanism of morphological control, as well as the excellent degradation-resistance of Al nanorods, will have broad impacts for biochemical sensing, surface protection, and energy harvesting in harsh environment.

**Methods:**

**Nanorod growth:** Al nanorod arrays are grown onto cleaned Si{100} substrates using electron beam evaporation PVD at varied vacuum levels. First, Si{100} substrates (Nova Electronic Materials) are ultrasonically cleaned in acetone, ethanol, and de-ionized water (Millipore) and are subsequently placed onto a precision machined mount for glancing angle deposition at the top of the vacuum chamber. The vacuum chamber is a stainless steel tank that is approximately 40 cm tall and 25 cm in diameter and relies on a mechanical vacuum pump and a turbo-molecular pump. Source material, 99.99% Al (Kurt J. Lesker) is placed in a graphite liner in the electron beam source at the base of the vacuum chamber. The Al source material is pre-melted via the electron beam before samples are introduced to the chamber and allowed 15 minutes under the energetic beam to outgas.

For deposition at $1 \times 10^{-2}$ Pa, the high vacuum stage, the turbo-molecular pump, is engaged for approximately five minutes and the base pressure reaches $5 \times 10^{-3}$ Pa. Vacuum is measured using an ionization vacuum gauge. The electron beam is then engaged and the deposition rate is monitored and controlled at 1.0 nm/s via quartz crystal microbalance for a total nominal thickness of 500 nm. The thickness is measured perpendicular to the source flux, and represents that of a continuous film. Vacuum level is measured throughout deposition with an ionization gauge and remains near $1 \times 10^{-2}$ Pa.

For deposition at $1 \times 10^{-5}$ Pa, the chamber is allowed to remain under high vacuum pumping for 24 hours and reaches a base pressure of $1 \times 10^{-5}$ Pa. The substrate is baked out at a temperature of 625 K during the initial 24 hours. The substrate heating is then removed and the substrate is allowed to return to room temperature. To further improve the vacuum, the substrate is blocked from flux via a sample shutter and chromium (Cr) is deposited onto the chamber walls via the electron beam source. After the deposition of Cr, the base pressure is further improved to $5 \times 10^{-6}$ Pa. To further decrease the amount of O present in the system during deposition, the Al source material is preheated by the electron beam for one hour under the high vacuum condition. Three times, the Al is heated to liquid and held for 20 minutes and cooled. Immediately following this process, Cr is deposited to remove the oxygen released from the source. After three cycles, the base pressure is reduced to $1 \times 10^{-6}$ Pa and Al deposition begins. The deposition rate is again monitored and controlled to 1.0 nm/s before the shutter is removed to expose the substrate. The Al nanorods are again grown to a nominal thickness of 500 nm. Vacuum level is measured throughout the deposition with an ionization gauge and remains near $1 \times 10^{-5}$ Pa.

To reach a substrate temperature of 225 K, liquid nitrogen is flowed into the chamber directly behind the substrate and the temperature is measured with K-type thermocouple. The liquid nitrogen is allowed to cool the entire fixture for one hour prior to deposition. The liquid nitrogen is then allowed to boil off and liquid water is added - ice forms and then acts as a thermal buffer layer. The system is allowed to equilibrate to 225 K and liquid nitrogen is added periodically to maintain the temperature, within a range of 200 K to 250 K.

**Nanorod Characterization:** Immediately after growth completes the Al nanorod arrays are removed from the deposition chamber and characterized using SEM and TEM. For SEM, Al nanorods are imaged using a FEI Quanta 250 Field Emission Scanning Electron Microscope. TEM is performed with Al nanorods that are grown directly onto carbon coated TEM grids or with Al nanorods drop coated onto Formvar TEM grids. Briefly, Al nanorod coated Si wafers are placed into a beaker of DI water and vigorously sonicated for one minute. The nanorods break off of the substrate at the base and are suspended in the water. The nanorod-water solution is then drop coated onto the TEM grid and allowed to dry under a gentle flow of argon. Immediately after drying TEM samples are imaged using a FEI Technai operating at 120 KeV.

**Thermal Annealing:** Thermal annealing experiments are performed in air using a resistance heated tube furnace. The annealing temperature is reached before the sample is placed inside the furnace on an alumina crucible. Timing begins when the sample is placed into the furnace and ends when the sample is removed. TEM samples are annealed while attached to the substrate, and are subsequently removed via sonication and drop coated onto TEM grids.

**Surface Enhanced Raman Spectroscopy:** SERS is performed using a Renishaw System 2000 in ambient after samples are sensitized with an Di-tetrabutylammonium cis-bis(isothiocyanato)bis(2,2'-bipyridyl-4,4'-dicarboxylato)ruthenium(II) (N719) dye solution. First, the N719 solution is created by adding 2.9 mg of N719 powder (Sigma Aldrich Product Number 703214) to 25 ml of reagent grade ethanol (Sigma Aldrich) and stirring for 12 hours at room temperature. Al nanorod substrates are then placed into the solution bath and allowed to remain there at a bath temperature 300 K for one hour. Annealed Al nanorod arrays are allowed to fully cool to room temperature before being placed into the N719 solution. Samples are then promptly removed and rinsed five times with a stream of ethanol to remove any excess dye solution. The samples are allowed to dry under air for 30 minutes and are then tested.

A 514.5 nm laser is used at an operating power of 50 mWatt. The laser is focused via the Raman microscope (Renishaw) through a 100 X magnification objective. Data is acquired through scanning for 10 seconds. For the baseline comparison, the Si{100} substrate of 2 cm x 2 cm is drop coated with a 50 µl droplet of the $1.0 \times 10^{-3}$ M N719 and is allowed to dry in air.

**References:**

1. Campion, A. and Kambhampati, P. Surface-enhanced Raman scattering. Chem. Soc. Rev. **27**, 241 (1998).

2. Shanmukh, S. et al. Rapid and sensitive detection of respiratory virus molecular signatures using a silver nanorod array SERS substrate. Nano Lett. **6**, 2630-2636 (2006).

3. Chaney, S. et al. Aligned silver nanorod arrays produce high sensitivity surface-enhanced Raman spectroscopy substrates. Appl. Phys. Lett. **87**, 031908 (2005).

4. Liu, Y-J. et al. Silver nanorod array substrates fabricated by oblique angle deposition: morphological, optical, and SERS characterizations. J. Phys. Chem. C **114**, 8176-8183 (2010).

5. Orendorff, C. et al. Aspect ratio dependence on surface enhanced Raman scattering using silver and gold nanorod substrates. Phys. Chem. Chem. Phys. **8**, 165-170 (2006).

6. Nikoobakht, B. and El-Sayed, M. Surface-enhanced Raman scattering studies on aggregated gold nanorods. J. Phys. Chem. A. **107**, 3372-3378 (2003).

7. Nuntawong, N. et al. Shelf time effect on SERS effectiveness of silver nanorods prepared by OAD technique. Vacuum **88**, 23-27 (2013).

8. Stagon, S. and Huang, H. C. Airtight metallic sealing at room temperature under small mechanical pressure. Sci. Rep. **3**, 3066 (2013).

9. Li, X. et al. High temperature surface enhanced Raman spectroscopy for *in-situ* study of solid oxide fuel cell materials. Energy Environ. Sci. **7**, 306 (2014).

10. Niu, X. et al. Smallest metallic nanorods using physical vapor deposition. Phys. Rev. Lett. **110**, 136102 (2013).

11. Huang, H. C. A framework of growing crystalline nanorods. JOM **64**, 1253-1257 (2012).

12. Zhang, R. and Huang, H. C. Another kinetic mechanism of stabilizing multiple-layer surface steps. Appl. Phys. Lett. **98**, 221903 (2011).

13. Liu, S., Huang, H. C. and Woo, C. Schwoebel-Ehrlich barrier: from two to three dimensions. Appl. Phys. Lett. **80**, 3295 (2002).

14. Lee. S. and Huang, H. C. From covalent bonding to coalescence of metallic nanorods. Nanoscale Res. Lett. **6**, 559 (2011).

15. Xiang, S. and Huang, H. C. Ab initio determination of Ehrlich-Schwoebel barriers on Cu{111}. Appl. Phys. Lett. **92**, 101923 (2008).

16. Hong, J. et al. Manipulation of spin reorientation transition by oxygen surfactant growth: a combined theoretical and experimental approach. Phys. Rev. Lett. **92**, 147202 (2004).

17. Hwang, E. and Lee, J. Surfactant-assisted metallorganic CVD of (111)-oriented copper films with excellent surface smoothness. Electrochem. Solid-State Lett. **3**, 138-140 (2000).

18. Sakai, A. and Toru, T. Ge growth on Si using atomic hydrogen as a surfactant. Appl. Phys. Lett. **64**, 52 (1994).

19. Gao, P., Ding, Y., and Wang, Z. Crystallographic orientation-aligned ZnO nanorods grown by a tin catalyst. Nano Lett. **3**, 1315-1320 (2003).

20. Jana, N., Gearheart, L., and Murphy, C. Wet chemical synthesis of high aspect ratio cylindrical gold nanorods. J. Phys. Chem. B **105**, 4065-4067 (2001).

21. Johnson, C. et al. Growth and form of gold nanorods prepared by seed-mediated surfactant-direct synthesis. J. Mater. Chem. **12**, 1765-1770 (2002).

22. Yata, M., Rouch, H. and Nakamura, K. Kinetics of oxygen surfactant in Cu(001) homoepitaxial growth. Phys. Rev. B **56**, 10579- 10584 (1997).

23. Au, M. et al. Free standing aluminum nanostructures as anodes for Li-ion rechargeable batteries. J. Power Sources **195**, 3333-3337 (2010).

24. Wang, P. et al. Size control of Cu nanorods through oxygen-mediated growth and low temperature sintering. Nanotechnology **20**, 085605 (2009).

25. Qiu, Z. et al. Raman spectroscopic investigation on TiO2-N719 dye interfaces using Ag@TiO2 nanoparticles and potential correlation strategies. Phys. Chem. Chem. Phys. **14**, 2217-2224 (2013).

26. Chu, H., Huang, Y. and Zhao, Y. Silver nanorod arrays as a surface-enhanced Raman scattering substrate for foodborne pathogenic bacteria detection. Appl. Spectrosc. **62**, 922-931 (2008).

27. Li, J. et al. Shell-isolated nanoparticle-enhanced Raman spectroscopy. Nature **464**, 392-395 (2010).

**Acknowledgements:** The authors acknowledge financial support from the Department of Energy Office of Basic Energy Sciences (DE-FG02-09ER46562).

**Author Contributions:** SPS and HH conceptualized the design and prepared the manuscript, and SPS carried out the growth and characterization of Al nanorods. XT and YL carried out the SERS characterization, and participated in the manuscript revision.

**Additional Information:** Supplemental material is linked to this article online.

**Competing Financial Interests:** The authors declare no competing financial interests.